\documentclass{jfm}    
\usepackage{graphicx} \usepackage{epstopdf, epsfig}
\usepackage[usenames,dvipsnames]{color}

\newcommand\BV{Brunt-V\"ais\"al\"a frequency}      
           
\usepackage{lpic}

\shorttitle{Scaling laws for mixing in unforced weakly rotating stratified turbulence}
\shortauthor{A. Pouquet, D. Rosenberg, R. Marino and C. Herbert}
\title{Scaling laws for mixing and dissipation in unforced rotating stratified turbulence} 
\author{A. Pouquet$^{1,2}$, D. Rosenberg$^{3}$, R. Marino$^{4}$ and C. Herbert$^{5}$}
\affiliation{$^{1}$National Center for Atmospheric Research, P.O.~Box 3000, Boulder, CO, 80307, USA.\\
$^{2}$Atmospheric \& Space Physics Laboratory, University of Colorado, Boulder CO, 80309, USA.\\
$^3$duaner62@gmail.com. \\
$^{4}$Laboratoire de M\'ecanique des Fluides et d'Acoustique, CNRS, \'Ecole Centrale de Lyon, Universit\'e de Lyon, \'Ecully, 69134, FRANCE. \\
$^{5}$\'Ecole Normale Sup\'erieure, 46 All\'ee d'Italie, Lyon, F-69364, FRANCE.  } \date{\today}
   
\begin{document} \maketitle
\begin{abstract}
We present a model for the scaling of mixing in weakly rotating stratified flows characterized by their Rossby, Froude and Reynolds numbers $Ro, \ Fr$, $Re$. It is based on   {quasi-}equipartition between kinetic and potential modes, sub-dominant vertical velocity $w$, and  lessening of the energy transfer to small scales as measured by a dissipation efficiency $\beta= \epsilon_V/\epsilon_D$, with $\epsilon_V$ the kinetic energy dissipation 
  and $\epsilon_D=u_{rms}^3/L_{int}$ its dimensional expression, $w, u_{rms}$  the vertical and {\it rms} velocities,  and $L_{int}$ the integral scale. We determine the domains of validity of such laws {for} a large numerical study of the unforced Boussinesq equations {mostly} on grids of $1024^3$ points, with 
$Ro/Fr\ge 2.5$, and with {$1600 \le Re\approx 5.4\times 10^4$;}  the Prandtl number is one, initial conditions are {either} 
isotropic and at large scale for the velocity, and zero for the temperature $\theta$, {or in geostrophic balance.} 
Three regimes in Froude number, as for stratified flows, are observed: dominant waves, eddy-wave interactions and strong turbulence. A wave-turbulence balance for the transfer time {$\tau_{tr}=N\tau_{NL}^2$,} with $\tau_{NL}=L_{int}/u_{rms}$ the turn-over time and {$N$ the Brunt-V\"ais\"al\"a frequency,}
 leads to $\beta$ growing linearly with $Fr$ in the intermediate regime, with a saturation at {$\beta\approx 0.3$ or more, depending on initial conditions}
 for larger Froude numbers. The Ellison scale is also found to scale linearly with $Fr$. The flux Richardson number $R_f=B_f/[B_f+\epsilon_V]$, with $B_f=N\left<w\theta \right>$ the buoyancy flux, transitions for roughly the same parameter values as for $\beta$. These regimes {for the present study} 
are delimited by ${\cal R}_B=Re Fr^2 \approx 2$ and $R_B\approx 200$. With $\Gamma_f=R_f/[1-R_f]$ the mixing efficiency, putting together the three relationships of the model allows for the prediction of the scaling $\Gamma_f\sim Fr^{-2}\sim {\cal R}_B^{-1}$ in the low and intermediate regimes for high $Re$, whereas for higher Froude numbers, $\Gamma_f \sim {\cal R}_B^{-1/2}$, {a scaling already found in observations:} as  turbulence strengthens, $\beta \sim 1$, $w\approx u_{rms}$, and smaller buoyancy fluxes  altogether correspond to a decoupling of velocity and temperature fluctuations, {the latter becoming} passive. 
 \end{abstract}
 
 \section{Introduction}

Mixing, which takes place for a large domain of flow parameters in fully developed turbulence (FDT), in engineering and geophysical flows, has been analyzed extensively~\citep{peltier_03, dimotakis_05, ivey_08}. One central issue is that of mixing efficiency, 
{which can be defined in many ways (see \citet{mashayek_17} and references therein),  for example as the ratio in the momentum equation of the buoyancy flux to the rate of kinetic energy dissipation (see \S \ref{S:MIX} for more details).}
 For example,  in the ocean, it affects chemistry and plankton dynamics, as well as the global circulation and hence the climate on long time-scales \citep{ivey_08, mcwilliams_16}. The mean circulation of the ocean and atmosphere {is}
 modified by a combination of stratification and  dissipation, 
{but rotation can also play a role.}
 In the oceanic context,  data indicates an enhanced vertical mixing that can be compared with the \citet{osborn_80} model, stating that the adimensionalized scalar diffusivity is proportional to the turbulence intensity parameter,  ${\cal R}_I=\epsilon_V/[\nu N^2]$, with an efficiency $\Gamma_f =B_f/\epsilon_V \approx 0.2$ as soon as ${\cal R}_I \gtrsim 10$ (see also~\citet{lindborg_08, karimpour_15});
 {$\epsilon_V$ is the energy dissipation rate, $\nu$ the kinematic viscosity, $B_f$ the buoyancy flux, and $N$ is the \BV.}
  Unsurprisingly, though, there is evidence that this efficiency coefficient does depend on the intensity of the turbulence within a stratified flow~\citep{smyth_01, ivey_08}, and exhibits  temporal variability associated with secondary instabilities of Kelvin-Helmoltz rolls~\citep{mashayek_13}, as well as a variation with Prandtl number \citep{smyth_01}. Indeed, mixing efficiency is in part governed by secondary instabilities that develop three-dimensional eddies at high Reynolds numbers. These eddies can dominate the whole process irrespective of Froude number but in ways that may depend on other parameters of the flow, {\it e.g.}, in the presence or not of shear,  leading to a lack of monotonicity in its variation with the  Richardson number \citep{peltier_03}. Thus, a one-parameter modeling  of such flows based on stratification alone may be insufficient. 

The amount of kinetic energy available for dissipation at small scales in rotating stratified turbulence (RST) is a crucial quantity for  sub-grid scale parameterizations in oceanic and climate models, and depends on the amount of energy that is transferred to these small scales in the presence of inertia-gravity waves, through breaking at small scales of the large-scale quasi-geostrophic (QG) balance 
\citep{lelong-riley_1991, staquet_rev_02,  riley_03},
and a lowering of the Richardson number below some critical value.  Wave-turbulence interactions allow for coupling to the mean flow \citep{finnigan_99}; they have been measured in the stratosphere  as well as in the upper ocean and lead to vertical mixing and enhanced dissipation
 {(see {\it e.g.,} \citet{klymak_08, vanharen_16g}).}
Few studies have considered mixing in decaying rotating stratified flows. Lagrangian diffusion in RST is studied in \citet{cambon_04};  with $N/f$ taking the values of 0.1, 1 and 10, as well as 0 and $\infty$, a reduced  diffusivity is found in the vertical, but not in the horizontal. The role of $N/f$,  with $f$  twice the rotation rate,  as a governing parameter for the intensity of lateral mixing with geostrophic adjustment has also been stressed in \citet{lelong_05}. Experimentally, it was shown in \citet{praud_06} that structures develop an aspect ratio proportional to $f/N$ (see also \citet{waite_06, kurien_14} for direct numerical simulations for the forced case): in RST, there is a progressive shift from a vertical scale due  entirely to stratification, the buoyancy scale $L_B=u_{rms}/N$, to one corresponding to QG where rotation and stratification are in balance with pressure gradients. 
{Finally,  \citet{dritschel_15} studied the influence of $N/f$ on large-scale quasi-geostrophic balance; they found it weak, the flows remaining balanced throughout the studied parameter regime, although vertical velocity increases with $f/N$ for $Fr^2<< Ro<<1$.}

 It is in this context that we now analyze  {several}   sets of direct numerical simulations (DNS) of decaying RST,
 {motivated by  atmospheric and oceanic applications,} with  Reynolds numbers up to  $Re\approx  {1.85} \times 10^4$, an upper value comparable to that in the Mesosphere and Lower Thermosphere~\citep{liu_13}. 
 We show that three simple scaling laws for the potential to kinetic energy ratio, the vertical to kinetic energy ratio, and the effective dissipation coefficient together lead to the recovery of a well-known scaling for mixing as a function of Froude number at high Reynolds numbers, 
 {and that these scalings persist even when significantly different initial conditions--i.e., those in quasi--geostrophic balance-- are used.}

\section{Problem setting} \label{S:EQ}   

The Boussinesq equations, with  constant rotation and stable stratification are: 
 \begin{eqnarray} 
 \partial_t {\mathbf u}  +  \nabla p  + {\mathbf u} \cdot \nabla {\mathbf u}  
 &=& 
  f {\bf u} \times \hat z  - N \theta \hat z  +  \nu \delta {\mathbf u}  \ , \label{eq:mom}  \\
 \partial _t \theta\  +  {\mathbf u} \cdot \nabla \theta 
 &=& N w +  \kappa \delta \theta\  , \label{eq:temp}   \end{eqnarray} 
 with ${\bf u}=(u,v,w)$ the incompressible velocity field, $\nabla \cdot {\bf u}=0$, and $p$ the pressure; $\theta$ represents  temperature (or density) fluctuations, in units of a velocity, since we want to stress the energetics of these flows. These fluctuations are 
super-imposed on
on a stably stratified background with a linear vertical profile $\bar \theta(z)=\theta_0+z\partial_z\bar{\theta},\ \partial_z\bar{\theta}<0$. 
Introducing the buoyancy $b=N\theta$ 
{(note the choice of sign),}
 with point-wise vertical flux $b({\bf x}) w({\bf x})$, as well as the  integrated buoyancy flux
\begin{equation}  B_f=\left<N\theta w\right> \ , \label{eq:flux0} \end{equation}
one recovers the more standard form of the equations in terms of $b$. 
$N=\sqrt{-(g/\theta_0)\partial_z\bar{\theta}}$ is the \BV,  and $\nu=\kappa$ are the  viscosity and  thermal diffusivity.
The Boussinesq equations  are integrated using direct numerical simulations.
A cubic box of $n_p^3=1024^3$ points is used for {56 runs (see Table \ref{tab1}),}
 with a linear dimension $L_{box}=2\pi$, resulting in  wave numbers in the range $1\le k \le k_M=n_p/3$ using a standard 2/3 dealiasing rule. All length scales defined below are thus expressed in terms of the fundamental length $L_{box}=2\pi$, which of course can be rescaled to the physical problem when necessary. 
{Another smaller set of runs, at lower resolutions, is also analyzed (see Table \ref{tab2}).} 
The pseudo-spectral code we use, GHOST (Geophysical High-Order Suite for Turbulence), is parallelized in a hybrid fashion with both MPI and Open-MP \citep{hybrid_11} and demonstrates scalability to in excess of 130,000 cores. It includes many solvers for fluid and magnetohydrodynamic  turbulence, 
 and it  now also has the capability to simulate non-cubic geometry \citep{mininni_17c}. GHOST has been tested  in the purely stratified case against the 
 numerical results of \citet{kimura_96} for random initial conditions, and of \citet{riley_03} for the   Taylor-Green flow. 

{For the runs of Table \ref{tab1},}
the potential energy is initially zero, and initial conditions for 
the initial velocity,
{with a non-zero vertical component,}
are isotropic, random and centered on the large scales ($2\pi/L_0=k_0=2.5$);  energies are defined as: 
$$E_V= \frac{1}{2} \int \|\mathbf{u}(\mathbf{x})\|^2 d^3\mathbf{x} \ , \ \ \ 
E_P=  \frac{1}{2} \int |\theta(\mathbf{x})|^2 d^3\mathbf{x} \ , \ \ \ E_T= E_V+E_P \ ,$$
$E_T$ being the total energy.
 They  can also be written in terms of their respective isotropic Fourier  spectra, with $\int E_{V,P}(k)dk=E_{V,P}$. Similarly, the kinetic, potential and total rates of energy dissipation are: 
\begin{equation}
\epsilon_V= \nu \int \|{\boldsymbol{\omega}}(\mathbf{x})\|^2 d^3\mathbf{x} \ , \ \ \  
\epsilon_P= \kappa \int \|\nabla \theta(\mathbf{x})\|^2 d^3\mathbf{x} \ ,  \ \ \  \epsilon_T= \epsilon_V+\epsilon_P  \ . 
\label{eq:eps} \end{equation} 
Note that $\epsilon_V$ can be measured {relative to its}
 dimensional evaluation of kinetic energy dissipation for a fully turbulent flow as: 
 \begin{equation}
\beta \equiv \epsilon_V/\epsilon_D \ , \  \epsilon_D\equiv u_{rms}^3/L_{int}  \ , 
u_{rms} = [\left< |{\bf u}|^2 \right>]^{1/2}  \  , \ L_{int}=2\pi \frac{\int [E_V(k)/k]\ dk}{\int E_V(k)\ dk} \ ,
 \label{beta2}  \end {equation}  
where $L_{int}$ is the integral scale \citep{monin_79}. $\beta$ is a key parameter of the phenomenological and theoretical understanding of the interactions between waves and  turbulence \citep{zakharov_92}, as  also discussed in \S \ref{SS:beta}. 
  
The governing dimensionless parameters of the Boussinesq equations measure the  strength of nonlinear interactions relative to dissipation, rotation and stratification; they are the Reynolds (Re), Rossby (Ro) and Froude (Fr) numbers, defined as usual as: 
 \begin{equation}
 Re=\frac{u_{rms}L_{int}}{\nu} \ , \ Ro=\frac{u_{rms}}{fL_{int}} \ , \  Fr=\frac{u_{rms}}{NL_{int}} \ ,
 \label{param}  \end{equation}
 with the Prandtl number $Pr=\nu/\kappa$ taken equal to unity. 
{The buoyancy Reynolds number, the Richardson number, and  the turbulent intensity are defined as:}
  \begin{equation} 
  {\cal R}_B\equiv Re Fr^2 \ , \ \ Ri\equiv [N/\left<\partial_z u_\perp \right>]^2 \ , \ \ {\cal R}_I\equiv \epsilon_V/[\nu N^2] \ . 
  \label{eq-rb-ri2} \end{equation}
{$Ri$ is based on a shear time computed on vertical gradients of the horizontal wind, namely $\tau_{shear}=[\left< \partial_z u_\perp \right>]^{-1}$. }
All these parameters are discussed further in the Appendix,  \S \ref{SS:beta0}. 
\\
In the presence of stratification, a variety of  length-scales can be relevant \citep{thorpe_87, mater_14}, {\it e.g.}
\begin{equation}
 L_B=2\pi \sqrt{E_V}/N \ \ , \ \  L_{Ell}= 2\pi\sqrt{E_P}/N \ \ , \ \ \ell_{Oz}=2\pi \sqrt{\epsilon_V/N^3} \ , 
\label{eq-ell} \end{equation}
or the buoyancy, Ellison  and 
{Ozmidov}
 scales. In purely stratified flows, $L_B$ is the scale for which the vertical Froude number becomes of order one \citep{billant_01}; it measures the thickness of the vertical layers. On the other hand, the  Ellison scale corresponds to the vertical distance traveled by a particle of fluid before being completely mixed, and it is thought to be significantly smaller than the integral scale in strongly stratified flows, 
 {as we shall show later (see Figs. \ref{f:LE}b).}
  $L_B$ and $L_{Ell}$  vary as $1/N$, but differ by a $\sqrt{E_V/E_P}$ field-amplitude ratio. 
 {Finally, the Ozmidov scale is the scale beyond which isotropy is thought to be recovered together with a classical Kolmogorov range.}

 \section{Global behavior and scaling} \label{S:STRUC}   
\subsection{Overview of the runs} \label{SS:TABLES}

 Runs with an emphasis on realistic parameters  for the mesosphere and lower thermosphere, and that overlap with the present data base, were investigated for the energy partition between waves and slow modes and the link with kinetic-potential energy exchanges in \citet{marino_15w}, as well as for parametric characteristic time-scale variations in \citet{rosenberg_16}
 {(see also \citet{rosenberg_17}).}
  Here, the runs 
 {on grids of $1024^3$ points,}
 cover the following parameter ranges
 {(see Table \ref{tab1}):}
   $0.11 \le Ro \le 41$, $1985 \le Re \le 18590$, $0.001 \le Fr \le 5.5$, $0.02 < {\cal R}_B < 1.2 \times 10^5$ and $2.47 \le N/f \le 312$.  
   {Two} purely stratified runs are included as well. Note that, even though the ratio in values of Reynolds numbers  across all these runs is close to ten, most runs are within a factor $\approx 4$ of each other in $Re$, 
 {with as high a value as can be realized on the chosen grid, thus breaking large-scale balance toward isotropization, 
 as studied already in \citet{herring_80} using a closure model of turbulence 
 {(see \citet{pumir_16, rubinstein_17} and \citet{ iyer_17} for recent references).}
 
 {Two other small series of runs at lower resolutions have been performed (see Table \ref{tab2}). The first study ({\bf Q} runs) is focused on the role of initial conditions, taking now  geostrophically balanced fields at $t=0$, which should radiate waves much less initially.
 The second small set of ({\bf Z}) runs  deals with the variation of effective dissipation $\beta$ defined in equation (\ref{beta2}) with Reynolds number at fixed $N/f=5$, with $Re$ varying by a factor in excess of 10, between $1650$ and $18590$, when including runs of Table \ref{tab1}.}
 
 All statistics are computed dynamically around the peak of dissipation, when the flow is most developed and starts its self-similar temporal decay. This is in contrast to what is done in \citet{stretch_10b}, where the data for mixing is taken when more than 90\% of the energy has dissipated,  after roughly ten turnover times.  Specifically, our data is averaged on a number of outputs around the peak of enstrophy, covering a relative variation in the amplitude of enstrophy of $\approx  2.5$\%. This results in using of the order of 18 outputs on average for each run, with no more than 49 and no fewer than 6; the physical time interval on which these averages were performed is of the order of a fraction of a turn-over time.
 We find that  $u_{rms}$ and $L_{int}$  do not vary much across the  {first large} 
 parametric study, from 0.66 to 0.89 for the former, and from 1.39 to 2.78 for the latter. 
Note that some of the runs tabulated in \citet{rosenberg_16} have been removed from the data set in Table \ref{tab1}, which has been reduced {from 65}
to 56 runs. This is because of various factors: archiving issues in view of the large data base that was created 
{several } years ago, or because the variation of enstrophy at peak was insufficient to satisfy the averaging criterion, or because some of the data files were corrupted. 

The accuracy of the computations is quantified through the ratio of the maximum  to the Kolmogorov dissipation wavenumber, $k_M/k_\eta$, 
{with $k_\eta=[\epsilon_V/\nu^3]^{1/4}$; this is done}
under the assumption that the small scales have recovered a  Kolmogorov spectrum, {\it i.e.} that the Ozmidov length scale is resolved, or for ${\cal R}_B \ge 1$,
 {which is the case for the majority of our runs.}}  For all flows {of Table \ref{tab1},} 
 we have $0.39 \le k_\eta/k_M \le 1.3$, with roughly 17\% slightly under-resolved runs which all have $N/f\ge 10$. We have also checked that the overall shape of the curves plotted in the figures did not depend on the resolution.

Strong activity develops at  small scales, with layer destabilization, as found as well by a number of authors in the purely stratified case. An example of such structures is given in the visualizations found in \citet{rosenberg_15} for a flow which corresponds rather closely to some of the runs computed in this data base 
{(specifically, run Id=11, 19, 33 and 43),}
but done on a grid of $4096^3$ points, allowing for a substantially higher Reynolds number. Prominent in this flow with $N/f=4.95$, $Fr\approx 0.0242$, $Re\approx 54000$, is the juxtaposition of large-scale eddies and an intense vortical activity at their rims in what can be called vortex lanes. Such a complex small-scale flow  corresponds to local overturning instabilities, with local Richardson numbers well below 0.25
 {for a substantial portion of the flow (see {\it e.g.} Figs. 10 and 11 in \citet{rosenberg_15}).}

\begin{table} \caption{\label{tab1}
{Nomenclature of the runs performed on grids of $1024^3$ points, with Id, $Fr, Ro$ and $Re$ the identification of runs and their Froude, Rossby and Reynolds numbers, the runs being ordered by $Fr$ (see also \citet{rosenberg_16}). Initial conditions are centered on the large scales and are isotropic for the velocity field, and zero for the temperature. Runs 62 to 65 are purely stratified. 
Note that the 9 runs marked with a star are not included in the present study (see \S \ref{SS:TABLES});  they are specifically runs Id=4, 8, 10, 21, 30, 38, 50, 62 and 63. }
 }  \begin{tabular}{ccccccccccccccc}   \hline
{\bf Id} &$Fr$ & $Ro$ & $Re$ & || & {\bf Id} & $Fr$ & $Ro$ &$Re$ & || & {\bf Id} & $Fr$ &$Ro$& $Re$ &    \\   \hline

{\bf 1}& 0.0013 & 0.129 & 10905 &  & {\bf 2} & 0.0023 &0.115  & 9895 &  & {\bf 3} & 0.0061 & 0.120 & 10680 &    \\
{\bf 4*}& 0.0064 & 0.633 & 9270   & . & {\bf 5} & 0.0073 & 0.225 & 13945 & . & {\bf 6}& 0.0116 & 0.305 & 14680 &   \\
{\bf 7}& 0.0119 & 2.98 & 13500   & . & {\bf 8*} & 0.0127 & 0.635 & 8930 & . & {\bf 9} & 0.021 & 0.147 & 11080 &  \\
{\bf 10*}& 0.0215& 0.464 & 13450 & . & {\bf 11} & 0.022&  0.116 & 10980 & . & {\bf 12} & 0.0262 & 4.58 & 10980&   \\
\hline

{\bf 13}& 0.028 & 0.14 & 10720 & . & {\bf 14} & 0.030 & 9.4 & 10520 & . & {\bf 15} & 0.033 & 4.58 & 13020 &   \\
{\bf 16} & 0.036 & 9.1 & 13200 & . & {\bf 17} & 0.038 & 0.140  & 10530 & . & {\bf 18} & 0.041 & 0.607 & 9840 &   \\
{\bf 19} & 0.042 & 0.211 & 14840 & . & {\bf 20} & 0.045 &  3.02 & 12790 & . & {\bf 21*} & 0.047 & 9.23 & 8880 &  \\
{\bf 22} & 0.048 & 4.5& 12370 & . & {\bf 23} & 0.049 & 4.57 & 18590 & . & {\bf 24} & 0.049 & 9.2 & 18550 &  \\
\hline

{\bf 25} & 0.049 & 9.3 & 12770 & . & {\bf 26} & 0.057 & 0.28 & 13730 & . & {\bf 27} & 0.057 & 0.14 & 9750 &  \\
{\bf 28} & 0.061 & 3.03 & 11650 & . & {\bf 29} & 0.062 & 0.61 & 9640 & . & {\bf 30*} & 0.067 & 920 & 13750 &  \\
{\bf 31} & 0.067 & 9.2 & 11730 & . & {\bf 32} & 0.073 &  3.04& 12210 & . & {\bf 33} & 0.086 & 0.43 & 12110 &  \\
{\bf 34} & 0.088 & 0.49 & 8525 & . & {\bf 35} & 0.09 &  4.6& 11010 & . & {\bf 36} & 0.10 & 9.3& 7720 &  \\
\hline

{\bf 37} & 0.10 & 6.9 & 8200 & . & {\bf 38*} & 0.10 & 0.49 & 8200 & . & {\bf 39} & 0.10 & 9.4 & 10750 &  \\
{\bf 40} & 0.10 & 7.0 & 10750 & . & {\bf 41} & 0.10& 7.1 & 16230 & . & {\bf 42} & 0.13 & 0.33& 7560 &  \\
{\bf 43} & 0.14 & 0.67 & 7600 & . & {\bf 44} & 0.14 & 0.98 & 7440 & . & {\bf 45} & 0.14 & 1.4 & 7330 &  \\
{\bf 46} & 0.16 & 9.8 & 9580 & . & {\bf 47} & 0.19 & 38 & 2520 & . & {\bf 48} & 0.20 & 5.0& 8720 &  \\
\hline

{\bf 49} & 0.20 & 10.1 & 8760 & . & {\bf 50*} & 0.21 & 41 & 6270 & . & {\bf 51} & 0.26 & 10.3& 8580 &  \\
{\bf 52} & 0.34 & 0.84 & 5020 & . & {\bf 53} & 0.38 &11.4  & 7120 & . & {\bf 54} & 0.40 & 42.2 & 1980 &  \\
{\bf 55} & 0.47 & 1.2 & 4500 & . & {\bf 56} & 0.55 & 11.1 & 7140 & . & {\bf 57} & 0.6 & 1.5 & 5010 &  \\
{\bf 58} & 0.89 & 2.2 & 4710 & . & {\bf 59} & 1.25 & 12.5 & 6470 & . & {\bf 60} & 2.7 & 12.5 & 4020 &  \\
\hline

{\bf 61} & 5.5 & 13.7 & 4260 & . & . & . & . & . & . & {\bf 62*} & 0.012 & $\infty$ & 15225 &  \\
{\bf 63*} & 0.027 & $\infty$ & 11800 & . & {\bf 64} & 0.07 & $\infty$ & 11490 & . & {\bf 65} & 0.20 & $\infty$& 8800 &  \\
 \end{tabular}  \end{table}

{In all the figures in the present work, different symbols are used for different binning, mainly in Rossby number; stars/asterisks are used for runs with QG initial conditions (ICs), whereas the hollow shapes are always for the $\theta(t=0)=0$ ICs. Furthermore, the sizes of all symbols refer to the resolution and Reynolds number (see caption of Fig. \ref{f:R1}). }

\subsection{Energy ratios} \label{SS:ENER}

\begin{table} \caption{ \label{tab2} 
{
Parameters for two other sets of DNS  identified by Id and  ordered, for each set, by their Froude number $Fr$, with 
$Ro$ and $Re$ the Rossby and Reynolds numbers computed at the time of maximum of enstrophy for each run. ${\cal R}_B=ReFr^2$, and $n_p^3$ is the total number of grid points for each run. 
In runs {\bf Zx}, $x=[1,8]$, the  initial conditions are large-scale isotropic and random for the velocity field,  and zero for the temperature as for the runs of Table \ref{tab1}, whereas in the {\bf Qx} runs, $x=[9,17]$, initial conditions are in geostrophic balance for velocity and temperature fluctuations.
For both sets, $N/f\approx 5$, a value close to what is found in the ocean. 
}}
\begin{tabular}{cccccccccccccc}  \hline
{\bf Id} & $n_p$  & $Fr$ & $Ro$ & $Re$ & ${\cal R}_B$  &  .& {\bf Id}& $n_p$ & $Fr$ &$Ro$& $Re$ & ${\cal R}_B$ &     \\
\hline

{\bf Z1}     &    256   &    0.042   &    0.208   &   3458  & 6.1    &  .  &   {\bf Z2}     &    512   &    0.063   &    0.316   &   6202  & 24.6  \\
{\bf Z3}     &    256   &    0.064   &    0.321   &   3358  & 13.7  & .   &   {\bf Z4}     &    512   &    0.064   &    0.321   &   6643  & 27.2 \\
{\bf Z5}     &    128   &    0.065   &    0.325   &   1694  & 7.1    & . &  {\bf Z6}        &    256   &    0.097   &    0.487   &   2951  & 27.8 \\
{\bf Z7}     &    256   &    0.651   &    3.255   &   1657  &  702  & . & {\bf Z8}     &    256   &    3.296   &   16.481   &   1706  & 18578 \\
\hline \hline

{\bf Q9}        &    256   &    0.007   &    0.036   &   5221  & .25  & . & {\bf Q10}        &    256   &    0.015   &    0.073   &   4973  & 1.1 \\
{\bf Q11}  &    256   &    0.039   &    0.197   &   3706  & 5.6 & . & {\bf  Q12}        &    128   &    0.067   &    0.335   &   1617  &  7.2  \\

{\bf Q13}        &    256   &    0.075   &    0.373   &   3130  & 17.6  & . & {\bf Q14}        &    512   &    0.076   &    0.382   &   6278  &  36.3  \\
{\bf Q15}       &    256   &    0.111   &    0.555   &   2537  &  31.2  & . & {\bf Q16}       &    256   &    0.577   &    2.817   &   2003  & 667 \\
 {\bf Q17}        &    256   &    1.290   &    6.451   &   2008  &  3341  & . &  \\

\end{tabular}  \end{table}

\begin{figure*}
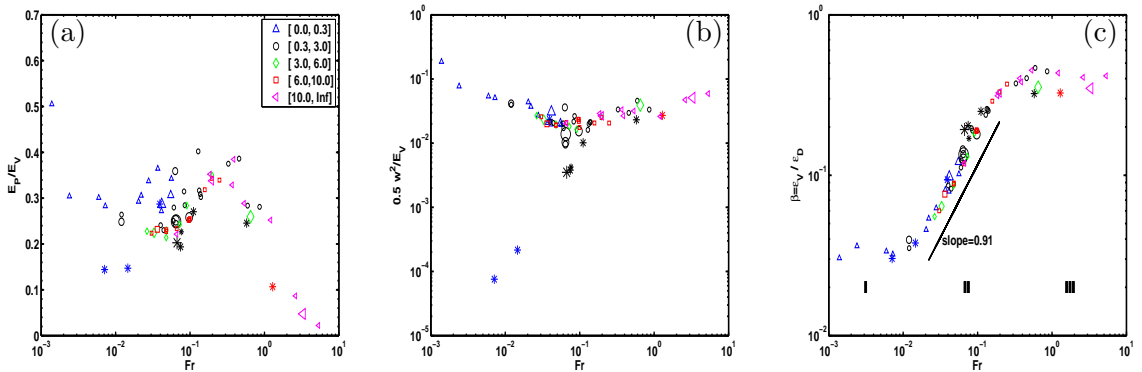
   
\vspace{-1.95truecm}
\hspace{-1.1cm} \begin{minipage}{0.27\textwidth}
\large \begin{lpic}[]{pouquet-figure-1a(0.25,0.34)}
\lbl[l]{37,197;(a)} \normalsize \end{lpic} \end{minipage} 
\hspace{1.4cm}\begin{minipage}{0.27\textwidth}
\large \begin{lpic}[]{pouquet-figure-1b(0.25,0.34)}
\lbl[l]{167,197;(b)}
\normalsize   \end{lpic}  \end{minipage} \vspace{-1.95truecm}
\hspace{1.4cm} \begin{minipage}{0.27\textwidth}
\large \begin{lpic}[]{pouquet-figure-1c(0.25,0.34)}
\lbl[l]{167,197;(c)} 
\normalsize \end{lpic} \end{minipage}
 \caption{(Color online) 
Variation with Froude number of the ratio of  potential to kinetic energy  ({\it  a}),  of vertical to kinetic energy ({\it b}), and of the rate of kinetic energy dissipation compared to its dimensional evaluation, 
$\beta=\epsilon_V/\epsilon_D$ ({\it c}). Colors/symbols indicate binning in Rossby number for all runs: blue triangles for $0< Ro \le 0.3$, black circles for $0.3< Ro \le 2.9$, green diamonds for $2.9< Ro \le 6.0$, red squares for $6.0< Ro \le 10.0$, and magenta inverted triangles for $Ro > 10$ (see insert). 
{The runs of Table \ref{tab2} are indicated either by a star for those with quasi-geostrophic initial conditions (Q runs), or by a hollow symbol for the runs with $\theta(t=0)=0$; their relative size is proportional to viscosity, thus  inversely proportional to $Re$ and to numerical resolution. We } 
 define the three dynamical regimes as: I  for strong waves ($Fr \lesssim 0.01$), II for eddy-wave interactions ($0.01 \lesssim Fr \lesssim 0.2$), and III for strong stratified turbulence ($Fr \gtrsim 0.2$), {as indicated in (c); note the quasi-linear scaling of $\beta$ with $Fr$ in regime II, 
namely $\beta \sim Fr^{0.91}$. }
}  \label{f:R1}    \end{figure*}

We show in Fig. \ref{f:R1} the  variation of central energetic quantities, as a function of Froude number, with binning in Rossby number. 
The widths of the bins are chosen so as to have approximately the same number of data points in each bin, as for all other figures.
The energy ratio $r_E\equiv E_P/E_V$ (left) varies roughly between  0.2 and 0.4, as long as the velocity and temperature remain coupled through buoyancy, i.e. for $Fr<1$. At high $Fr$, $E_P/E_V$ becomes negligible since the velocity  is no longer constrained effectively by the waves and we have a quasi passive scalar regime with $\theta(t=0)=0$ 
{for most of the runs.} The scaling in $Fr^{-2}$ at high $Fr$, as advocated for oceanic turbulence in \citet{wells_10} (see also \citet{maffioli_16l} for purely stratified flows), may be present as well, although we have a scarcity of points in that domain. We thus conclude that 
{in the intermediate regime of wave-vortex interactions:} 
\begin{equation} 
\theta_{rms}\sim u_{rms} \ .      \label{lawT} \end{equation}

The result $r_E \approx 1$ from below (see also \citet{mater_13}) is compatible with the experimental, observational and numerical data compiled in \citet{zilitinkevich_08}. It is also compatible with $\epsilon_P/\epsilon_V\approx 1/3$, as measured in the stratosphere \citep{lindborg_06}.
In fact, such a quasi-equipartition of energy is found in a large range of wave-numbers, as shown in \citet{marino_15w} in the forced case.
It immediately implies that the Ellison scale $L_{Ell}$ goes as $L_{int} Fr$, for which in fact the scaling is excellent (see Fig. \ref{f:LE}b below), and that
 $L_{Ell}\sim L_B\sim u_{rms}/N$. Note also that a pattern is  discernible in $E_P/E_V$ with, on average, higher relative vertical velocity and higher relative potential energy at low Rossby number.  As a function of Reynolds number, $r_E$ decreases on average for $Re$ larger than $\approx 10^4$, because of the 
 initial conditions  (not shown).

The ratio of vertical to total kinetic energy 
{$\left<w^2/2 \right>/E_V$} 
is given in  Fig.~\ref{f:R1}(b). At low $Fr, {\cal R}_B$,  with weak nonlinear mode coupling, its high value 
{for the runs of Table \ref{tab1}}
is due to initial conditions, taken with a rough equipartition between velocity modes in all directions, in order to let anisotropy develop dynamically.
 A similar reduction in vertical velocity for rapidly rotating flows in the absence of stratification was observed in laboratory experiments \citep{vanbokhoven_09}, together with a weaker  dissipation. 
In an intermediate range of parameters, around $Fr\approx 0.1$, there is a small  plateau,  the stratification being strong enough to prevent most of the vertical motions; so,
\begin{equation}  
{w_{rms} \lessapprox u_{rms} \ .}
 \label{lawW} \end{equation}
 As turbulence strengthens with increasing $Fr$, vertical motions develop slowly 
 {after that plateau, with an approximate scaling}
 $\left< w^2 \right>\sim \left< u_\perp^2 \right> Fr^{1/4}$. The origin of such a weak scaling is not clear, and no scaling is found in terms of $N/f$.
 We recall here that the vertical velocity $w\hat e_z$ is also a direct measure of wave activity since, in a wave-vortex decomposition as performed {\it e.g.} in \citet{bartello_95, herbert_16}, vortical modes have vanishing $w$.
The presence of rotation facilitates vertical motions in the form of upward propagating inertial waves along Taylor columns that would form if there was  no stratification, as clearly observed,  including when the small scales develop strong vorticity \citep{davidson_06, mininni_12}: the vertical velocity is leaving a trace of the influence of rotation on the system and its increase is consistent with a wave-vortex analysis. 
The scaling in equation (\ref{lawW}) may seem unexpected. Dimensional analysis  using incompressibility at large scale would predict $w_{rms}/u_{rms} \sim Fr$, ruled out by the data
{with isotropic ICs, but in agreement with the data with QG initial conditions at low Froude numbers, with this energy ratio reaching the value obtained with the isotropic ICs for $Fr\approx 0.1$.} We know that strong vertical velocities develop for intermediate $Fr$ when the buoyancy and nonlinear terms balance each other, leading to a ``saturation'' vertical energy spectrum $E_\parallel \sim N^2 k_\parallel^{-3}$. The  model developed in \citet{rorai_14}, adding the buoyancy flux to a Vieillefosse description of intermittency for FDT, leads to $E_\parallel \approx E_P\approx E_V$ under the hypothesis that the characteristic vertical scale is the buoyancy scale. In these intermittent  regions identified by low Richardson numbers, high vertical velocities 
{appear,} due to strong turbulent stirring leading to overturning of layers.
Note that it is argued in \citet{maffioli_16d} that gradients are much larger in the vertical, or that $w$ is really at small scale. Thus, these authors advocate  
$\left< w^2 \right>\sim \epsilon_V/N$. With $\epsilon_V\sim  \epsilon_D Fr$ (see \S \ref{SS:beta} {below),} this leads  {again}
 to $w/u_\perp\sim Fr$. On the other hand, in their paper, $\beta$ is viewed as a constant denoted $A_k<1$, independent of dimensionless parameters, so their estimate is rather $w/u_\perp\sim Fr^{1/2}$,   unlike our data {at high $Fr$.}

\subsection{Effective {\it versus} dimensional dissipation and the three regimes of RST} \label{SS:beta}

The dissipation efficiency of rotating stratified flows $\beta$ 
is shown in Fig. \ref{f:R1}(c) as a function of $Fr$; it clearly displays three regimes. For small $Fr$ up to $Fr\approx 0.01$, $\epsilon_V$ is low and constant. Similar low dissipation efficiency, of the order of a few percents, is obtained for runs corresponding to the Upper Troposphere and Lower Stratosphere  region, with low Froude numbers, as analyzed for example in \citet{paoli_14} using a sub-grid model. 
 We find that, above $Fr\approx 0.01$, $\beta$  grows quasi-linearly with $Fr$, with a least-square fit giving a slope of {0.91} 
  after which $\beta$  saturates, for $Fr\gtrsim 0.2$. Thus,
\begin{equation}
\beta= \epsilon_V/ \epsilon_D \sim Fr \   \ \ \ \ \ \ \ \ [{\it Intermediate \ regime, II]} \ , 
 \label{lawE} \end{equation} 
 thereby defining the three dynamical regimes of RST, 
 {namely}
 I, II \& III, ordered by increasing $Fr$.
Such a scaling
 is also found when examining  the small-scale energy flux in the forced case in the presence of an inverse cascade
 \citep{marino_15p}. $\beta$ saturates at a value close to unity for highly turbulent flows at higher Froude numbers, as found as well in \citet{maffioli_16d}, although their values at peak of enstrophy are a bit higher. 
 It may be related to the fact that they compute in boxes of small aspect ratio, between $1/4$ and $1/6$, a geometry that 
{can favor}  vertical gradients and shear, {and which can lead} to a more active turbulence, {as found in} \citet{mininni_17c}.  
 When the turbulence strengthens, so does the direct energy cascade through baroclinic instability, frontogenesis and nonlinear coupling of eddies  \citep{mcwilliams_16}.
 {It should finally be noted that we do not expect actual transitional values of $\beta$, between regimes I \& II, and  II \& III, to be similar for similar control parameters in different flow geometries, but we do expect the scaling $\beta\sim Fr$ to hold, as shown here contrasting isotropic {\it versus} QG initial conditions (see also the dimensional argument in the Appendix showing how the transfer time to small scales is moderated by the stratification).}

\begin{figure*}
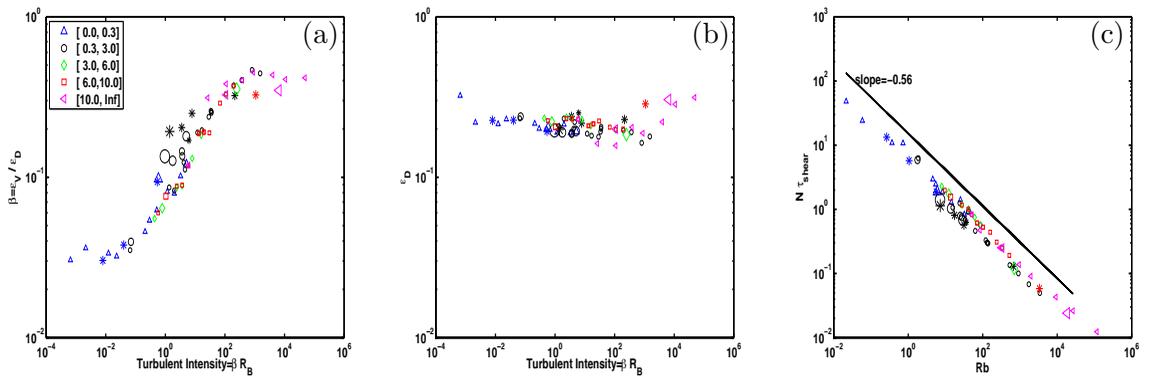
   
\vspace{-1.95truecm}
\hspace{-1.1cm} \begin{minipage}{0.27\textwidth}
\large \begin{lpic}[]{pouquet-figure-2a(0.25,0.34)}
\lbl[l]{168,197;(a)}
\normalsize \end{lpic} \end{minipage}
\hspace{1.4cm}\begin{minipage}{0.27\textwidth}
\large \begin{lpic}[]{pouquet-figure-2b(0.25,0.34)}
\lbl[l]{168,197;(b)}
\normalsize   \end{lpic}  \end{minipage} \vspace{-1.95truecm}
\hspace{1.4cm} \begin{minipage}{0.27\textwidth}
\large \begin{lpic}[]{pouquet-figure-2c(0.25,0.34)}
\lbl[l]{168,197;(c)}
\normalsize \end{lpic} \end{minipage}
 \caption{(Color online) 
Variation of $\beta$ ({\it a}) and $\epsilon_D=u_{rms}^3/L_{int}$ ({\it b}) with turbulent intensity parameter ${\cal R}_I=\epsilon_V/[\nu N^2]$ . In (c) is  shown $Ri^{1/2}\equiv N\tau_{shear}$ as a function of the buoyancy Reynolds number ${\cal R}_B=\beta^{-1} {\cal R}_I$, 
with $\tau_{shear}= \left<\partial_z u_\perp \right>$. All plots have binning  in Rossby number, 
{and symbols are as described in the caption of Fig. \ref{f:R1}.}
  The scaling 
{$N \tau_{shear}\approx {\cal R}_B^{-0.56}$} 
extends through  regimes I \& II with lower  ${\cal R}_B$ and $Fr$, 
 {and possibly regime III.}
}  \label{f:R2}    \end{figure*}
 
If we now examine the variations of $\beta$ with ${\cal R}_I=\epsilon_V/[\nu N^2]$, we  see  in Fig. \ref{f:R2}(a) that 
{we again have good scaling throughout, with} 
 $\beta\sim {\cal R}_I^{1/3}\sim \beta^{1/3}{\cal R}_B^{1/3}$. It is easy to show that this is compatible with  equation (\ref{lawE}), since  $\beta\sim {\cal R}_B^{1/2}\sim Fr$, omitting a dependency in $Re^{1/3}$,
 {although it appears clearly in Fig. \ref{f:R2} that, at fixed ${\cal R}_I$ and lower $Re$, $\beta$ is measurably larger since $Fr$ is larger.}
This  indicates that care must be taken when interpreting data as a function of dimensionless parameters.
In ${\cal R}_B$, the transitions between the three regimes occur respectively for $\approx 2$ and $\approx 200$. Such values 
are relevant for example in the ocean thermocline, where ${\cal R}_B\approx  10-100$  \citep{fleury_94}, as well as in lakes in which an average for ${\cal R}_B$ is $\approx 200$ \citep{bouffard_13}.

The dimensional kinetic energy dissipation $\epsilon_D$ (Fig. \ref{f:R2}b) is constant across parameters, 
{and across initial conditions,}
except for very small or large $Fr$ values. 
{A slight trend towards smaller values at higher ${\cal R}_I$, which can be attributed to smaller {\it rms} velocities, is discernible.}
Finally, we show in  Fig. \ref{f:R2}(c) that $Ri^{1/2} \sim {\cal R}_B^{-1/2}$, a scaling compatible with $Ri\sim Fr^{-2}$ at constant Reynolds number.
At low $Fr$, this relation gives the strength of vertical gradients (slanted because of rotation), and at higher $Fr$, it indicates a progressive return to isotropy and to only a single time-scale determining the dynamics, transfer and dissipation of such turbulent flows. 
Note that the three results in equations (\ref{lawT}--\ref{lawE}) may not be entirely new but, taken together, they define the key ingredients for establishing the scaling of the mixing efficiency which is discussed in \S \ref{S:MIX}.

\section{Mixing and dissipation} \label{S:MIX}
 \subsection{Definitions of mixing efficiency and flux Richardson number}

Irreversible mixing is  found in the laboratory to be triggered by merging Kelvin-Helmoltz billows \citep{patterson_06}, highly unstable as $Re$ increases. 
In the absence of rotation, parameter space 
{has been} separated into three regions in terms of $Fr$ and $Re$ \citep{luketina_89}: for small Re and Fr, waves are dominant and there are no turbulent motions, whereas for high Fr and Re, isotropic turbulence prevails. The intermediate region with roughly $Fr\le 1,\ {\cal R}_B \ge 10$ is where turbulence is anisotropic and strongly interacting with waves. The data on which these conclusions are based comes from the analysis of turbulent plumes active in tidal estuary flows~\citep{stillinger_83, stacey_99}. 
We find similar transitions with mild rotation, as shown in \S \ref{SS:beta}.

 In terms of  {the temporal evolution of}  vertical kinetic and potential energy density, one is led to compare the buoyancy flux $B_f=N\left< w \theta \right>$ with the  dissipation rates,  the Coriolis force  not affecting the   energy balance. Performing space-averaging, one can write:  
$$D_tE_V= -B_f + \epsilon_V \ , \ \ \ D_tE_P= B_f + \epsilon_P \ .  $$
 In order to quantify the relative magnitudes of these terms, several expressions have been introduced in the literature. Concerning the momentum equation,  one traditionally defines the flux Richardson number $R_f$ and its associated mixing efficiency $\Gamma_f$  as:   
\begin{equation}
R_{f}=  \frac{B_f}{B_f+\epsilon_V} \ \ ,  \ \ \Gamma_f=\frac{R_{f}}{1-R_{f}}=\frac{B_f}{\epsilon_V} \ .      \label{flux1} \end{equation}

\begin{figure*}
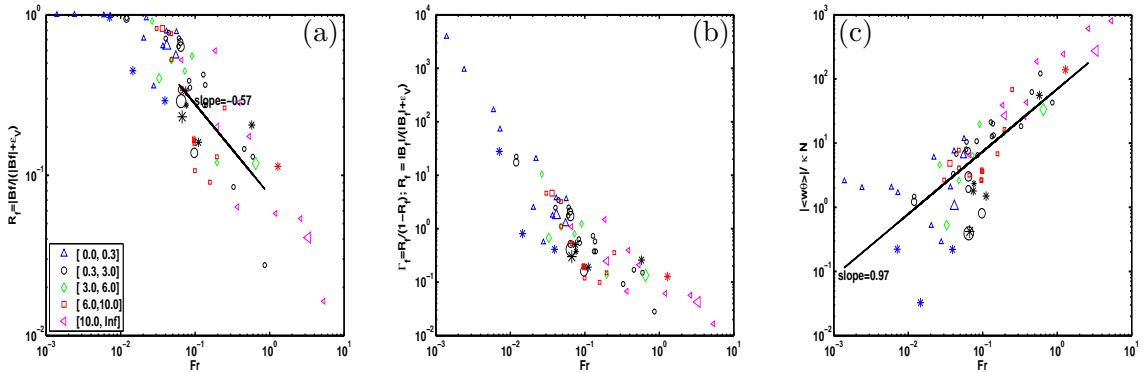
   
\vspace{-1.95truecm}
\hspace{-1.1cm} \begin{minipage}{0.27\textwidth}
\large \begin{lpic}[]{pouquet-figure-3a(0.25,0.34)}
\lbl[l]{168,197;(a)} \normalsize \end{lpic} \end{minipage} 
\hspace{1.4cm}\begin{minipage}{0.27\textwidth}
\large \begin{lpic}[]{pouquet-figure-3b(0.25,0.34)}
\lbl[l]{168,197;(b)} \normalsize   \end{lpic}  \end{minipage} \vspace{-1.95truecm}
\hspace{1.4cm} \begin{minipage}{0.27\textwidth}
\large \begin{lpic}[]{pouquet-figure-3c(0.25,0.34)}
\lbl[l]{35,197;(c)}
\normalsize \end{lpic} \end{minipage} 
 \caption{(Color online) 
Variation with Froude number of the flux Richardson number $R_f$ and  
of the mixing efficiency $\Gamma_f$ ({\it a,b}), 
{both defined in equation \ref{flux1},}
as well as of the effective diffusivity $\kappa_{\rho}/\kappa$ ({\it  c}), with 
$\kappa_\rho=N^{-1}\left< w \theta \right>$. Binning is performed in $Ro$
{(see Fig. \ref{f:R1} for symbols).}
 A transition at $Fr\approx 0.02$ is   seen in all three plots.  {Scalings are given as indications.}
}  \label{f:S1}    \end{figure*}

 The functional variation of $R_f$ with gradient Richardson number is central to numerical studies of geophysical flows. 
The mixing efficiency $\Gamma_f$  is singular for $R_f=1$, {\it i.e.} for fully mixed potential and kinetic modes (see \citet{mashayek_13, salehipour_15, mashayek_17} for a discussion on the definitions of mixing efficiency). This corresponds to negligible kinetic energy dissipation, {\it i.e.} a limit of zero Froude number. As we shall see in Fig. \ref{f:S1}, $\Gamma_f$ does reach high values at low $Fr$, in excess of $10^3$.
 Many recent works indicate  variations with parameters as $Re$  grows. 

\begin{figure*}
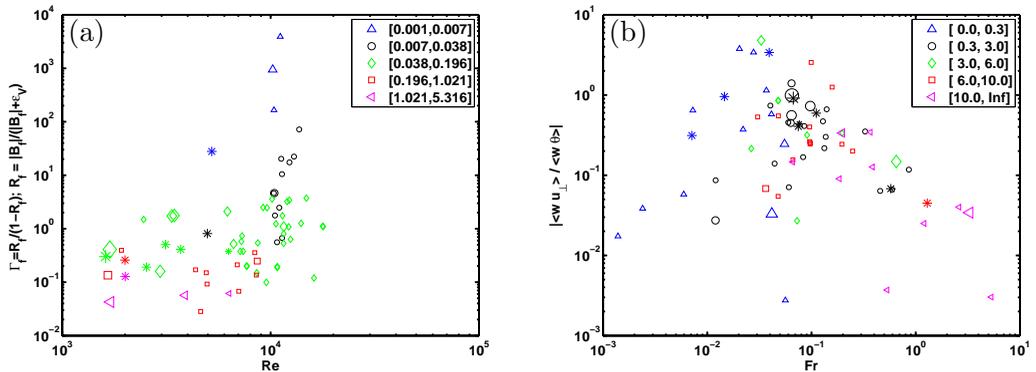
   
\vspace{-1.95truecm}
 \hspace{-0.7cm}  \begin{minipage}{0.47\textwidth}
\large \begin{lpic}[]{pouquet-figure-4a(0.35,0.34)}   
\lbl[l]{35,197;(a)} \normalsize \end{lpic} \end{minipage}
\hspace{0.7cm}\begin{minipage}{0.47\textwidth}
 \large \begin{lpic}[]{pouquet-figure-4b(0.35,0.34)} \lbl[l]{35,197;(b)}  
\normalsize   \end{lpic}  \end{minipage} \vspace{-1.95truecm}
 \caption{(Color online) 
{\it (a):} Variation of mixing efficiency $\Gamma_f$ with Reynolds number, 
{with binning in Froude number (see insert).} 
{\it (b):} Variation with $Fr$  of the ratio of the vertical flux of horizontal velocity to the vertical flux of buoyancy, 
{with binning in Rossby number. The size of symbols for both plots is described in the caption of Fig. \ref{f:R1}.}
}  \label{f:S2}    \end{figure*}

 \subsection{Mixing and effective diffusivity as a function of parameters}

We evaluate $\Gamma_f$ at  peak of dissipation, whereas in \citet{stretch_10b}, it is computed as the ratio of two time integrals after more than $90\%$ of the energy has dissipated. Since energy decay  in  turbulence is self-similar, 
these two methods should lead to comparable scalings. 
We show in Figs. \ref{f:S1}(a,b) $R_f$ and $\Gamma_f$ as a function of Froude number. At low $Fr$, $\epsilon_V$ is negligible compared to $B_f$, and $R_f\approx 1$.   A sharp transition occurs for $Fr\approx 0.02$, with $R_f$  decreasing continuously thereafter, and no visible saturation. The decline in $R_f$ begins in the intermediate range in which waves and vortices interact strongly. Similarly, $\Gamma_f$ shows a transition for $Fr\approx 0.02$, with a change in slope from $\Gamma_f\approx Fr^{-2}$ to $\approx Fr^{-1}$, as $Fr$ grows (approximate scalings), and with a variation of several orders of magnitude. At higher $Fr$ (regimes II \& III), a power-law scaling seems likely, as for $R_f$. 
 The variation of $R_f$ with Richardson number mirrors its variation in terms of $Fr$ (not shown). Another example of strong variation of $\Gamma_f$ is found in~\citet{bluteau_13} where $\Gamma_f=0.2$ is only valid in the range $7 < {\cal R}_B < 100$,  {whereas} we find it for  $100\lesssim {\cal R}_B\lesssim 1000$. 

The overall variation of $\Gamma_f$ with $Fr$ and ${\cal R}_B$ is similar to that found in the absence of rotation but with shear \citep{mater_14}. These authors further note that the centroid of such a curve  depends on what flows are  studied,  as for example in the data of \citet{lozovatsky_13} where the centroid is shifted to higher ${\cal R}_B$. This presents a challenge, since  parameterization schemes are mostly based on DNS, which may still be at too low a value of ${\cal R}_B$, and since using ${\cal R}_B$ implies studying variations in terms of both stratification (through $Fr$) and of turbulence (through $Re$). 
{The variations with Reynolds and buoyancy Reynolds numbers are discussed further in \S \ref{S:NEW}.}

In Fig. \ref{f:S1}(c) we plot the effective diffusivity $\kappa_\rho$, relative to the  molecular diffusivity $\kappa$. It is proportional to the buoyancy flux $B_f$.
Taking the notation in \citet{ivey_08}:
\begin{equation}
\kappa_{\rho}= B_f/N^2 \ \ , \ \ \kappa_{\rho}/\kappa = \left<w\theta \right>/[N\kappa]  \ ;
\label{eq:turb} \end{equation}
$\kappa_\rho$ is comparable to $\kappa$ at low $Fr$, and its increase with $Fr$ is close to a linear variation, a least-square fit giving 
 {$\kappa_{\rho}/\kappa \approx Fr^{0.97}$.}
Finally, a saturation begins to occur at high $Fr$. This behavior can be interpreted as being due to an  increase in buoyancy flux because of more vigorous stirring when ${\cal R}_B$ increases, at relatively constant $Re$.
When comparing the high-$Fr$ values of $\kappa_\rho/\kappa$ to the model proposed in \citet{barry_01}, a rough agreement is obtained.
Similarly, when examining a compilation of observational data for both salinity and temperature in the ocean together with numerical data for purely stratified flows, it is found in \citet{bouffard_13} that a transition occurs in $\kappa_\rho$, at ${\cal R}_B\approx 100$, between a ${\cal R}_B^{3/2}$ and a ${\cal R}_B^{1/2}$ scaling (see also \citet{shih_05} for the latter).
This enhancement of dissipation and  of transport coefficients, such as anomalous diffusivity, is expected in turbulent flows as shown in numerous studies of FDT~\citep{ishihara_09}, as well as in the strongly stratified case~\citep{ivey_08}. In this latter instance, it is probably due to strong intermittency, a signature of strongly stably stratified flows \citep{rorai_14}, such as in the planetary boundary layer \citep{finnigan_99}, or in the ocean~\citep{dasaro_11}. 

We show in Fig. \ref{f:S2}(a) the mixing efficiency $\Gamma_f$ as a function of Reynolds number.
 Around $Re\approx 10^4$, which covers many of the runs of Table \ref{tab1}, $\Gamma_f$ takes on a variety of values corresponding to how strongly the flow is stratified, with variations in excess of 1000.
 {However, at lower $Re$ (for most of the runs of Table \ref{tab2}), $\Gamma_f$ remains at lower values, the peak in $\Gamma_f$ being linked to the development of small-scale instabilities as $Re$ and ${\cal R}_B$ grow.}
Finally, another measure of the small-scale mixing efficiency of a flow is the ratio of the vertical fluxes of horizontal velocity to that of temperature fluctuations, as analyzed in~\citet{zilitinkevich_13}. It is  shown in Fig. \ref{f:S2}(b) as a function of $Fr$; no clear scaling emerges although one could advocate a $Fr^{-1}$ decrease for regimes II and III. Also, this ratio appears to be higher in the intermediate regime, on average.
    
\begin{figure*}
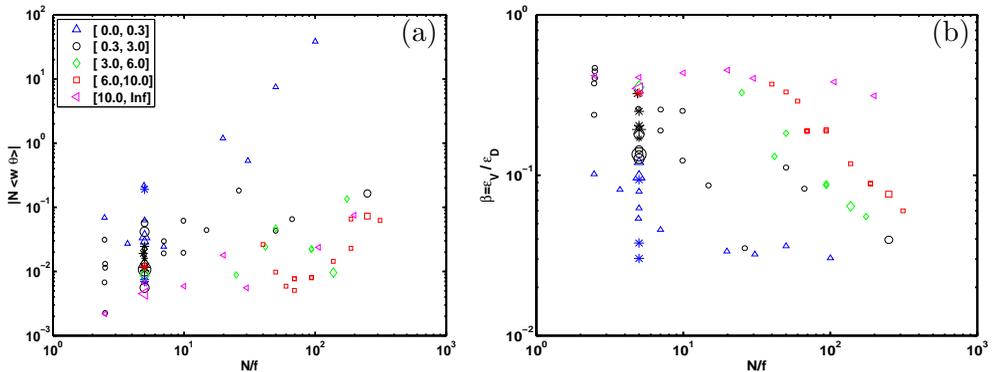
   
\vspace{-1.95truecm} \begin{minipage}{0.47\textwidth} 
\large \begin{lpic}[]{pouquet-figure-5a(0.33,0.34)} 
\lbl[l]{172,197;(a)}
\normalsize \end{lpic} \end{minipage}
 \hspace{-0.3cm} \begin{minipage}{0.47\textwidth}
\large \begin{lpic}[]{pouquet-figure-5b(0.37,0.34)}
\lbl[l]{174,197;(b)} \normalsize   \end{lpic}  \end{minipage} \vspace{-1.95truecm} 
 \caption{(Color online) 
Variation with  $N/f=Ro/Fr$ of the buoyancy flux $B_f=\left< Nw \theta \right>$ ({\it a}), and of the dissipation efficiency 
$\beta=\epsilon_V/\epsilon_D$ ({\it b}), with  binning in $Ro$ (see insert);
{symbols are as in Fig. \ref{f:R1}. For a given $N/f$, both can take a large range of values.}
}  \label{f:rot}    \end{figure*}

\section{The combined roles of rotation and stratification} \label{S:ROT}

{The addition of rotation} leads to the propagation of inertia-gravity waves whose dispersion relation 
depends on $N/f$. Thus, the atmosphere, with $N/f\sim 100$, and the ocean where $N/f\lesssim 10$  may differ in their statistical properties.
For {all runs of this paper,}  $N/f\ge 2.5$, so that stratification dominates. It is thus not surprising that the classical picture of mixing in stratified flows has not been changed in a significant way {when weak rotation is included but} in the absence of scale separation and of forcing.
We do see an effect of rotation on the magnitude of the potential energy (see {\it e.g.} Fig. \ref{f:R1}a), strong rotation  altering the large scales where the  energy is contained. It was shown in \citet{marino_15p} that rotation and stratification play complementary roles in the relative strength of the direct and inverse 
{constant-flux} energy cascades in the forced case: the  small-scale cascade is weaker the smaller the Froude number, and conversely the  large-scale  cascade is stronger the smaller $Ro$ is, in both cases affecting the effective dissipation of energy in the small scales and thus, presumably, the mixing properties of such flows. For the flows of Table \ref{tab1}, all micro Rossby numbers,  $R_\omega=\omega_{rms}/f$, are  larger than $3.1$, with $Ro\ge 0.11$.
Thus, the small-scale vorticity created by the nonlinear dynamics of the flow, 
{including in the presence of strong waves,}
is dominant at small scale, compared to the imposed rotation; note that such values for $R_\omega$ are plausible for geophysical flows.  

Figs. \ref{f:rot}(a,b) give the variations with $N/f$ of the buoyancy flux $B_f$ and of {the dissipation efficiency} $\beta$.
At a given $N/f$, there is less buoyancy flux the higher the Rossby number, and for a given bin in $Ro$, $B_f$ is larger the higher $N/f$: as the Froude number increases, the buoyancy flux decreases. The {efficiency}
 of dissipation is higher the higher the Rossby number (Fig. \ref{f:rot}b), again at fixed $N/f$. Perhaps the scaling of $\beta$ with $Fr$ is somewhat modified by rotation, leading to the slightly sub-linear law observed in Fig. \ref{f:R1}(c): there may be less direct energy transfer for strong rotation, as is observed in the forced case \citep{marino_15p}, and thus it takes a higher Froude number to reach a given level of dissipation. 
A clear effect of the Rossby number on dissipation in RST was also shown in the forced case in \citet{pouquet_17p}. 

Thus, the absence in the overall statistical properties of clear scaling in $Ro$ 
simply shows that the energy transfer to the small scales is dominated by a combination of stratification and turbulence. 
However, the primary purpose of this study is not to examine the role of rotation directly. Such a study would be helped by analyzing flows with $N/f<1$, a parameter range which is  relevant neither to the atmosphere nor to the ocean. For $N/f>1$, the most important role of rotation in such flows is 
 the  triggering of an inverse cascade of energy to the large scales,   attenuating the energy transfer to the small scales \citep{pouquet_13b, marino_15p, pouquet_17p}. In the absence of forcing and of scale separation, such an inverse cascade cannot develop in general.

\section{The Ellison scale and a generalized mixing efficiency} \label{S:DISC}

\begin{figure*}
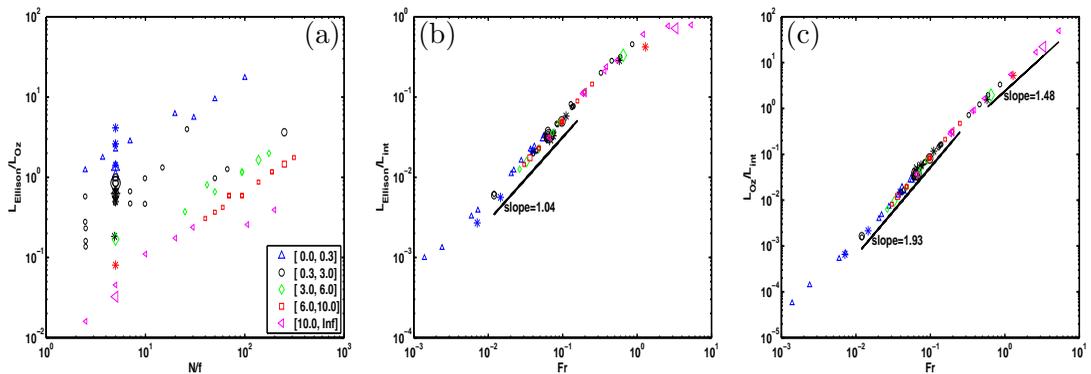
   \vspace{-1.95truecm}
\hspace{-1.1cm} \begin{minipage}{0.27\textwidth} 
\large \begin{lpic}[]{pouquet-figure-6a(0.25,0.34)}
\lbl[l]{169,197;(a)} \normalsize \end{lpic} \end{minipage}
 \hspace{1.cm} 
\begin{minipage}{0.27\textwidth}
\large \begin{lpic}[]{pouquet-figure-6b(0.25,0.34)}
\lbl[l]{35,197;(b)}  \normalsize \end{lpic} \end{minipage}
 \hspace{1.cm} \begin{minipage}{0.27\textwidth}
\large \begin{lpic}[]{pouquet-figure-6c(0.25,0.34)}
\lbl[l]{35,197;(c)} \normalsize   \end{lpic}  \end{minipage} \vspace{-0.05truecm}
\caption{(Color online) 
Ellison scale $L_{Ell}=\theta_{rms}/N$ relative to the Ozmidov scale $\ell_{Oz}$ ({\it a}) or the integral scale $L_{int}$ ({\it b}), as a function of $N/f$  ({\it a}), and of $Fr$ ({\it b)}. In {\it (c)} is given $\ell_{Oz}/L_{int}$ as a function of $Fr$.
All plots have binning in $Ro$
{(and see caption of Fig. \ref{f:R1}).}
Note the small dispersion in the  scaling of length scales versus Froude number
{(approximate scalings are given).}
} \label{f:LE} \end{figure*}

Similarly to characteristic time scales, one can also examine characteristic length scales. A comparison of the Thorpe and Ellison  scales in stratified turbulence was performed in \citet{mater_13} (see also \citet{dillon_82}). The Thorpe  scale $L_T$ corresponds to the vertical distance a parcel of fluid must be moved to produce a stable density profile, suppressing inversions, and as such gives an idea of the size of local mixing structures in the fluid; it was  computed as a function of ${\cal R}_B$ for the purely stratified case in \citet{metais_89}. In \citet{mater_13}, 
$L_T$ is found to be strongly linearly correlated with the Ellison scale $L_{Ell}$. Furthermore, the Thorpe length normalized by the Ozmidov scale is found to vary as $Fr^{-3/2}$ for $Fr>1$ and as $Fr^{-1/2}$ for $Fr<1$, the latter with an excellent scaling for $0.05 \le Fr \le 0.3$. Thus, $L_{T} \sim L_B$ for $Fr<1$, whereas $L_{T} \sim L_{int}$ for $Fr>1$. In the latter case, stratification is weak and structures in density follow the integral length scale, whereas in the former case of strong stratification, density changes occur over the vertical layer width, i.e. the buoyancy scale. 
Note that this also implies that, at small Froude number, $L_{T}/L_{int}\sim Fr$  and that $E_P\sim E_V$ \citep{mater_13}. 

In Fig. \ref{f:LE} are shown the variations of $L_{Ell}$ normalized by the Ozmidov scale (a) and  integral  scale (b), as a function of $N/f$ (a), and of $Fr$ (b),  with binning in Rossby number.  We conclude that the Ellison scale is larger, the larger $Fr$ is, as expected.
 As a function of Froude number, $L_{Ell}/\ell_{Oz}$ decreases, in a linear fashion for the intermediate regime, and with little dispersion among the runs 
 {(not shown).}
 One could argue that with $\ell_{Oz}= [\epsilon_V/N^3]^{1/2}$, for  small Froude number, $\ell_{Oz}/L_{int}\sim Fr^{2}$ in the intermediate regime in which 
 $\beta\sim Fr$, whereas for high $Fr$, $\ell_{Oz}/L_{int}\sim Fr^{3/2}$, 
 {in rough agreement with scaling laws, }  
 as indicated in Fig. \ref{f:LE}(c) giving $\ell_{Oz}/L_{int}=f(Fr)$ with least-square fits of respectively $1.93$ and  {$1.48$,}
 the transitions  taking place for $Fr\approx 0.01$ and $Fr\approx 0.2$, in agreement with the transitions for $\beta$ (see Fig. \ref{f:R1}c). 

On the other hand, the linear variation $L_{Ell}/L_{int}\sim Fr$ in Fig. \ref{f:LE}(b), with  a least-square fit giving {$\sim Fr^{1.04}$,}
  is a direct consequence of the scaling law $\theta_{rms}\sim u_{rms}$. The only transition in this power-law behavior takes place for $Fr={\cal O}(1)$, in which case $L_{Ell}\lesssim L_{int}$, with $L_{Ell}$ remaining smaller than $L_{int}$ because of the $1/k$ factor in the definition of $L_{int}$. 
There is also an indication of a slight saturation at low $Fr$. 

\begin{figure*}
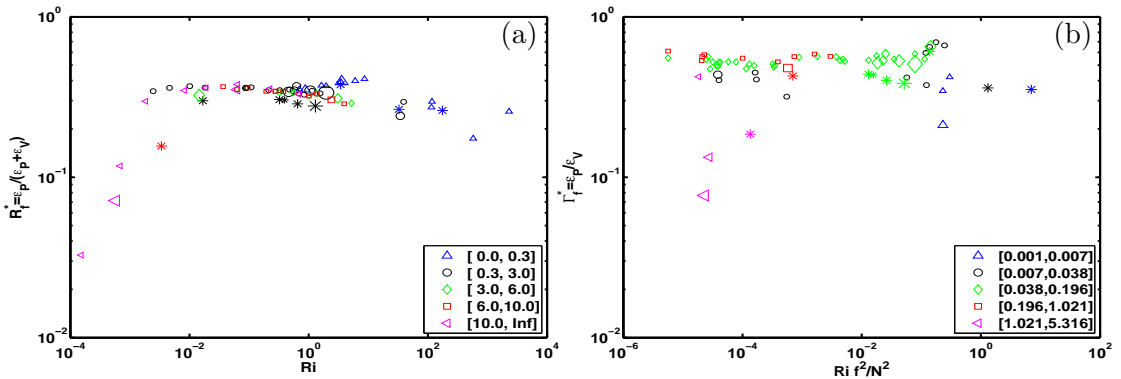
   \vspace{-1.95truecm}
\hspace{-1.1cm} \begin{minipage}{0.43\textwidth}
\large \begin{lpic}[]{pouquet-figure-7a(0.4,0.34)} 
\lbl[l]{175,197;(a)} \normalsize \end{lpic} \end{minipage} 
\hspace{1.4cm}\begin{minipage}{0.43\textwidth}
\large \begin{lpic}[]{pouquet-figure-7b(0.4,0.34)} 
\lbl[l]{175,197;(b)}  \normalsize   \end{lpic}  \end{minipage} \vspace{-1.95truecm} 
\caption{(Color online) 
{Potential energy dissipation normalized by total dissipation,}
 $R_f^\ast=\epsilon_P/\epsilon_T$, {\it vs.} Richardson number $Ri$ ({\it a}), and 
  $\Gamma_f^\ast=\epsilon_P/\epsilon_V$ {\it vs.} $Rif^2/N^2=[\tau_{shear} f]^2$  {\it  (b)}.
Binning is performed in $Ro$ in (a) and in $Fr$ in (b) (see inserts, and caption of Fig. \ref{f:R1}).
} \label{f:last} \end{figure*}

Another measure of the relative importance of terms in the Boussinesq equations is defined through  the ratio of the two dissipative terms 
{for momentum and temperature,}
 which can differ even at unit Prandtl number. 
We use the following definitions (see also  \citet{osborn_80, venayagamoorthy_16}): 
 \begin{equation}  
R_f^\ast= \frac{\epsilon_P}{\epsilon_T} \  \ , \ \   \Gamma_f^\ast = \frac{R_f^\ast}{1-R_f^\ast}= \frac{\epsilon_P}{\epsilon_V} \  \ , 
  \label{flux2} \end{equation} 
with $\epsilon_T = \epsilon_V+\epsilon_P$ already defined in equation (\ref{eq:eps}). $\Gamma_f^\ast$, called the irreversible mixing efficiency in \citet{mater_14}, relates to the partition of energy between the kinetic and potential modes, {\it i.e.} to the importance of the waves versus nonlinear eddies
{at small scales.} 
$R_f^\ast$ is shown in Fig. \ref{f:last}(a) as a function of $Ri$. In the first regime of strong waves ($Ri>1$),  it is small since the influence of initial conditions 
 prevails, and similarly for the low $Ri$ regime in which the potential-kinetic exchanges are inefficient, leading to an  abrupt decrease
{in $R_f^\ast$.}
{For intermediate values of $Ri$, almost from $10^{-3}$ to roughly $10$, $R_f^\ast$ stays rather constant in a range between $0.3$ and $0.4$.}
{Note also that} this data is consistent with the variations of $\epsilon_{V}$ and $\epsilon_{P}$ with $Fr$  studied in \citet{sozza_15} in a thin layer box.

We finally examine in Fig. \ref{f:last}(b) the variations of $\Gamma_f^\ast$ with $Rif^2/N^2=[\tau_{shear}f]^2$, 
{thus}
combining the effects of stratification and rotation.
A rather small  variation of $\Gamma_\ast \approx 0.5$ is obtained, with the  low points corresponding to small $f$. There is  a slow decline for small $Ro$ and large $Ri$ which may be related to sensitivity to initial conditions at low $Fr$: in regime I, 
$\epsilon_V\approx \epsilon_P$ is compatible with $E_P\approx E_V$ since, when the waves are strong to moderate, there is little nonlinear transfer and the dissipation is mostly contained in the large scales. 
Also, the highest value of $\Gamma_f^\ast \approx 0.7$ for $Fr\approx 0.01$, corresponds obviously to flows with comparable kinetic and potential energy dissipation. This is likely associated  in that regime to strong waves and  intermittent bursts which are due to wave breaking which temporarily relax the flow to a quasi-equipartition of kinetic and potential energies across a wide range of  scales, the more so the smaller the scale, as observed for example in~\citet{rosenberg_15}. 
Our results corroborate those of \citet{venayagamoorthy_16}: different measures of mixing, such as $\Gamma_f$ or $\Gamma_f^\ast$, give rather equivalent information, although it is not clear if this result will persist in the presence of forcing, at much higher buoyancy Reynolds numbers.

{ \section{Role of Reynolds number} \label{S:NEW}}

\begin{figure*}
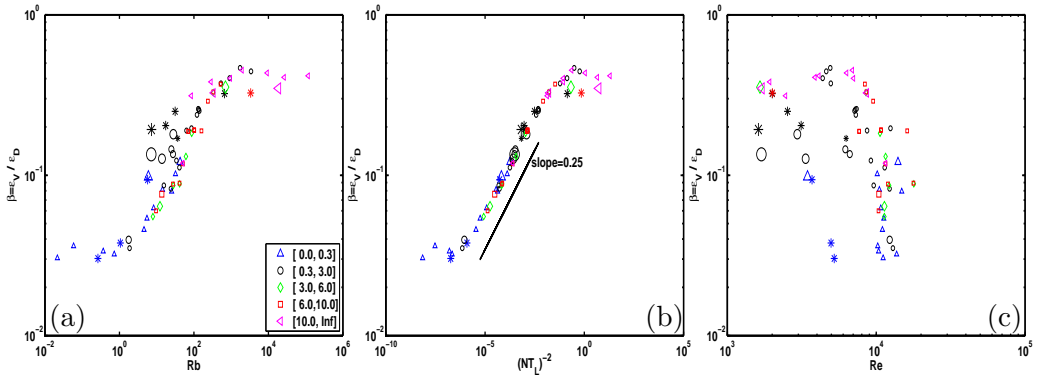
   
 \vspace{-1.95truecm}
\hspace{-1.05cm} \begin{minipage}{0.3\textwidth}
\large \begin{lpic}[]{pouquet-figure-8a(0.25,0.34)}
\lbl[l]{35,87;(a)}  \normalsize \end{lpic} \end{minipage}
\hspace{0.35cm}\begin{minipage}{0.3\textwidth}
\large \begin{lpic}[]{pouquet-figure-8b(0.25,0.34)}
\lbl[l]{170,87;(b)}   \normalsize   \end{lpic}  \end{minipage}  \vspace{-1.95truecm}
\hspace{0.25cm} \begin{minipage}{0.3\textwidth}
\large \begin{lpic}[]{pouquet-figure-8c(0.25,0.34)}
\lbl[l]{170,87;(c)}  \normalsize \end{lpic} \end{minipage}
 \caption{(Color online) 
{Dissipation efficiency $\beta$ as a function of (a) ${\cal R}_B$, (b) $[NT_L]^{-2}$, and (c) $Re$ 
(see \S \ref{SS:A} for a discussion of $NT_L=NE_V/\epsilon_V$). Note the scaling $\beta\sim [NT_L]^{-1/2}$ in (b).}
}  \label{f:new2}    \end{figure*}

{The variation of the intensity of the turbulence in rotating stratified flows can be measured by the Reynolds number, as well as by the buoyancy Reynolds number. Having high-enough $Re$ and ${\cal R}_B$ is known to be important for turbulent flows, to allow for coherent structures to develop including in the presence of strong stratification (see {\it e.g.} \citet{laval_03}). However, from a numerical point of view, having high $Re, {\cal R}_B$ for low $Fr$ is quite challenging and remains a goal for the near future. When taking the data for the runs of Tables \ref{tab1} and  \ref{tab2} for a possible scaling of the dissipation efficiency $\beta$ with ${\cal R}_B$, we observe some scatter (see Fig. \ref{f:new2}a). Specifically, 
we see that in the intermediate regime (${\cal R}_B$ between 10 and a few $100$), at fixed ${\cal R}_B$, there is a measurable variation in $\beta$, by contrast to regimes I, and to a lesser extent regime III. This scatter in regime II is larger than when examining variations with the Froude number itself, irrespective of the rotation (see Fig. \ref{f:R1}(c)). In Fig. \ref{f:new2}b, we see that overall, there is markedly less scatter when plotting $\beta$ as a function of the parameter 
$[NT_L]^{-2}$ as discussed in \citet{mater_14} (see the Appendix, \S \ref{SS:A}); $T_L\equiv E_V/\epsilon_V$ is the effective kinetic energy transfer time. Expressing $\epsilon_V=\beta \epsilon_D$, we see that $NT_L=[\beta Fr]^{-1}$; thus, the choice of the abscissa in Fig. \ref{f:new2}(b) is to be able to make a direct comparison with ${\cal R}_B$ which scales as $Fr^2$ at fixed $Re$. We also note that the regime transitions in terms of Richardson number occur for $Ri\approx 0.1$ and $Ri\approx 10$ (not shown).
 \\
The difference in data points scatter between Fig. \ref{f:new2}(a,b) can be attributed to the variations with Reynolds number, as displayed in  Figure \ref{f:new2}(c). It shows clearly that the Reynolds number alone does not allow for predicting the effectiveness of dissipation, and consequently that of mixing efficiency, with a wide scattering of data points for $\beta$, irrespective of the initial conditions tested in this paper. However, we note that at a given $Re$, QG initial conditions lead to a substantially lower dissipation efficiency. 
We further note that Reynolds numbers are still quite low for these runs, when comparing with geophysical flows. Similar conclusions can be drawn for the variation with $Re$ of the ratio of the Ellison scale normalized by the integral scale (not shown). 
}\vskip0.12truein

 \section{{Discussion,} conclusion and perspectives} \label{S:CONCLU}

 A  parametric study of  {mildly} rotating stratified turbulence without forcing leads to a rather {systematic}
  quantitative assessment of its mixing and dissipative properties which depend  on the Froude number provided the Reynolds number is high enough. 
  Three different regimes are observed, in agreement with previous studies of  purely stratified flows. 
  {These regimes are} also identifiable in terms of the interaction parameter ${\cal R}_I$. The three basic laws illustrated in Fig. \ref{f:R1} are compatible with an intermediate regime characterized by the dynamics of  waves and eddies interacting nonlinearly weakly, even though the full weak turbulence formalism leading to a set of closed integro-differential equations in terms of energy spectra cannot work for stratified flows because of a Froude number in the vertical of order unity \citep{billant_01}.   It is thus somewhat remarkable that the simple phenomenology embodied in the parameter $\beta\sim Fr$, {\it i.e.} the efficiency of the turbulent dissipation, based on a ratio of characteristic time scales (see equation (\ref{IK})), may still  apply on average for such flows. 

Together with $\theta_{rms}\sim u_{rms}$ and a scaling for $w/u_\perp$ going as a 
{quasi-}constant at intermediate $Fr$,
these laws imply   that the mixing efficiency $\Gamma_f\sim Fr^{-2}$ as soon as $Fr>0.01$, and  $\Gamma_f\sim Fr^{-1}\sim {\cal R}_B^{-1/2}$ for $Fr\lesssim {\cal O}(1)$. {We emphasize that the actual values of the control parameter for the change of regimes may depend on the geometry and topology of such flows.}
In the intermediate regime,  $\beta\sim Fr$, showing the connection between buoyancy flux  and nonlinear transfer leading to dissipation, with 
$\Gamma_f \beta^2 \sim 1$. Note that, with ${\cal R}_I\sim Fr {\cal R}_B$, this scaling law in the intermediate regime differs when expressed using ${\cal R}_I$.
Finally the mixing efficiency measured in terms of  the ratio of potential to kinetic 
 energy dissipation, is shown to be rather constant.
The scaling $\Gamma_f\sim Fr^{-1} \sim {\cal R}_B^{-1/2}$, in regime III at high $Fr$ and ${\cal R}_B$, simply stems from the decoupling of the velocity and temperature, together with $\beta\approx 1$, leading to $B_f\sim N$. 

Note also that $\Gamma_f$ and $R_f$ seem to be more sensitive to parameters with a clear indication of the three physical regimes in terms of $Fr$, ${\cal R}_B$ or ${\cal R}_I$, than either $\left< wu_\perp \right>/\left<w \theta \right>$ or $E_P/E_V$. {Furthermore,}
 if $w^2/u_\perp^2\sim Fr$ as advocated in \citet{maffioli_16d}, then the phenomenological arguments developed in our paper lead straightforwardly to $\Gamma_f\sim {\cal R}_B^{-3/4}$,  which cannot be entirely ruled out given the   scatter in data points for $\Gamma_f$, although it is not compatible with the data of Fig. \ref{f:R1}(b) with $w_{rms}/u_{rms}\sim 1$.
{We verified that taking as initial conditions geostrophically balanced flows did not alter our conclusions; similarly, having non-zero potential energy, but still unbalanced and with $w\not= 0$ ICs, we obtained the same fundamental scalings. } 

Local variations in Richardson number may trigger local density micro-structures, as observed in the ocean \citep{phillips_72, peltier_03}. If the agreement of our results, without shear but with rotation, with previous results mainly for sheared purely stratified flows is striking, it remains to be seen whether it will persist in the presence of forcing, i.e. in the presence of a strong inverse cascade.
{We note that}
\citet{waite_06} already observed  three regimes in the presence of forcing, with a switch for the energy cascade from  predominantly inverse to  direct.

Within the confines of the present parametric study with a wide range of buoyancy Reynolds numbers,  
$\Gamma_f$ is in fact rather variable, as in the purely stratified case. Rotation is essential for the existence of a dual constant-flux cascade of energy, implying two-dimensional (horizontal) lateral mixing as well as vertical.
If such mixing  occurs in proportion to the ratio of the inverse to direct cascade, it  will scale as $[Ro Fr]^{-1}$ \citep{marino_15p}.
 In the absence of forcing, with large-scale initial conditions and with  rotation weaker than stratification, all  effects associated with the presence of rotation are severely quenched. As discussed in \citet{mashayek_13}, shear can induce vortex pairing at moderate Reynolds number, reinforcing the potential for an inverse cascade in the presence of rotation but, on the other hand, as $Re$ increases, 3D instabilities take over and  shearing  leads  to  enhanced dissipation. 
 
 Another issue, when incorporating rotation or stratification, will be to consider the role of anisotropy on statistics, spectra and structures, the role of nonlocal interactions among scales, and the role of potential vorticity $P_V$ and the magnitude of its nonlinear part; these will be the topic of 
{future work.} Several other extensions of the present study are desirable. On the one hand, a {larger} 
scanning in terms of Reynolds numbers is needed, but only feasible today at values comparable to or lower than what is presently achieved in this paper, without using modeling such as eddy viscosity or hyper-viscosity, or some other partial truncation of modes such as computing in boxes with small aspect ratio. 
  From the numerical standpoint, the  condition $\ell_{Oz}>>\eta,\ {\cal R}_I>>1$ for strong and stratified turbulence to develop is hard to fulfill even with resolutions allowed by high-performance computing using available present-day technology.
  {For example,  in \citet{debruynkops_15}, the highest buoyancy Reynolds number that is reached, on a grid of $8192^2X4096$ points, is $\approx 220$, still quite low compared to atmospheric and oceanic values (see also \citep{iyer_17} for a FDT   run on a grid of $8192^3$ points with a Taylor Reynolds number of $1300$). } In the context of  the turbulent planetary boundary layer (see {\it e.g.}~\citet{sukoriansky_05}), 
one can write simplified expressions for vertical mixing, governed by vertical velocity, and horizontal mixing on the basis of a return to isotropy model; this leads to agreement between these models, and laboratory  and atmospheric data. One can also model the time-evolution of a mixing event by following characteristic length scales \citep{smyth_01}. 
 
 For the oceans, the collapsing of mixing efficiency at  high ${\cal R}_{I,B}$ embodied in the $\Gamma_f\sim {\cal R}_B^{-1/2}$ scaling, might imply the lessening of water mass motions in the ocean, by at least a factor of 2 as found for the Antarctic Bottom Water~\citep{delavergne_16}. 
 The tide impinging upon oceanic bottom topography leads to the formation of small scales which, beyond the Ozmidov scale, become isotropic. This could imply that the strong dissipation and mixing which is observed, for example at the Hawaian ridge \citep{klymak_08}, is propagating upward, in particular at mid latitudes where rotation plays a role, to scales of the order of 1 $km$. This may lead to exchanges of light and dense waters and abyssal sinking \citep{ferrari_16}, thereby affecting the net circulation patterns of the ocean. 
 Thus, a better understanding of the dynamics of rotating stratified turbulence, and of the scaling of its mixing and dissipative properties with  {control}
  parameters, may lead to better parametrization schemes to model more accurately these global phenomena.
\vskip0.2truein

{\it Acknowledgments: AP is thankful to LASP and Bob Ergun for support, and to Colm Caulfied for a useful discussion while at IPAM. RM acknowledges support from the {\it PRESTIGE} program coordinated by Campus France (co-financed under Marie Curie FP7 PCOFUND-GA-2013-609102) and the {\it PALSE} program at the University of Lyon. 
Computations were performed at the National Center for Atmospheric Research, 
{through an ASD allocation, as well as a new (2017) allocation of background time. NCAR}  is supported by the National Science Foundation.
{Finally, we also acknowledge  requests by the reviewers to perform more runs (see Table \ref{tab2}),
as well as to simplify the text.}}

\section{Appendices}  \label{S:APP}
{Many parameters and characteristic time-scales and length scales have been defined in the literature for rotating stratified turbulence, and we regroup some of them here for completeness. They allow for the definition of slightly different dimensionless parameters for which we also give an overview.}

\subsection{Appendix A: Characteristic time scales} \label{SS:A} 

{The four global control parameters of the Boussinesq equations written in \S \ref{S:EQ} can be written as the ratio of large-scale characteristic times, namely: 
\begin{equation}
 Re=\frac{\tau_{diss}}{\tau_{NL}}, \ Ro=\frac{\tau_{wr}}{\tau_{NL}}, \  Fr=\frac{\tau_{wg}}{\tau_{NL}} \ ,
 \label{paramt}  \end{equation}
with $\tau_{diss}=L_{int}^2/\nu, \ \tau_{NL}=L_{int}/u_{rms}$, $\tau_{wg}=1/N$ and $\tau_{wr}=1/f$ respectively the dissipation and  eddy turn-over times, and the gravity and inertial wave periods; finally, $Pr=\tau_\kappa/\tau_{diss}$ with $\tau_\kappa=L^2_{int}/\kappa$. The integral scale $L_{int}$ was defined in \S \ref{S:EQ}.
When linearizing the Boussinesq equations, one obtains inertia-gravity modes of frequency 
$\omega_k= \pm \sqrt{N^2k_\perp^2 + f^2 k_\parallel^2}/k$, with $k_{\parallel,\perp}$ referring to the vertical and horizontal directions (see {\it e.g.} \citet{bartello_95}).
However, for the sake of simplicity, we define the above parameters  using isotropy, {\it i.e.} omitting [$k_\perp, k_\parallel$] factors which would appear through the dispersion relation. One can define an effective transfer time for the kinetic energy, measured directly from observational or numerical data, as:
\begin{equation}
T_L\equiv E_V/\epsilon_V \ , \label{eq:TL} \end{equation}
 whereas $\tau_{NL}=E_V/\epsilon_D=\beta T_L$ is based on a-priori large-scale characteristics of the flow, with $\beta=\epsilon_V/\epsilon_D$ as defined in equation (\ref{beta2}).} 

{In the absence of  imposed shear, the Richardson number is based on a shear time $\tau_{shear}$ built from the vertical gradient of the mean horizontal wind $u_\perp$:
  \begin{equation}
\tau_{shear}=1/  \langle \partial_z u_\perp \rangle \ \ , \ \ Ri= [N \tau_{shear}]^2   .  \label{richard} \end{equation}
As such, $Ri$ can be viewed as measuring the strength, in terms of time-scales, of the formation of internal turbulent shear layers due to nonlinear interactions to that of the vertical layers due to the gravity waves, omitting the effect of rotation.}
 
{In the presence of several characteristic time and length scales, dimensional analysis is undetermined, even without rotation and for a unit Prandtl number.
 One particular set of parameters has been proposed in \citet{mater_14}, 
and we now analyze it, with some variation in notation. The crucial point is to emphasize the difference between the effective kinetic energy dissipation rate, $\epsilon_V$, and its dimensional evaluation, $\epsilon_D$, through their ratio $\beta=\epsilon_V/\epsilon_D$. \\
It is traditional in wave turbulence \citep{zakharov_92} to model the weaker transfer of energy to small scales due to the waves, when compared to a fully turbulent flow, by introducing a transfer time written {\it a priori} on dimensional grounds as:}

{
\begin{equation}
\tau_{tr}\equiv  \tau_{NL} \frac{\tau_{NL}}{\tau_{wg}} = \frac{\tau_{NL}}{Fr} \ ,
\label{IK} \end{equation}
using the small parameter adequate for the problem at small scale, here $Fr<<1$.
Thus, $\tau_{tr}> \tau_{NL}$, as expected. In the purely rotating case, one would use $\tau_{wr}=\tau_{NL}Ro$ 
\citep{cambon_89}.}

{It is then  deduced that consistency between the two definitions of a transfer time, namely taking $T_L$ and $\tau_{tr}$ to be proportional, immediately implies that one must have:
\begin{equation}
\beta \sim Fr 
\label{true} \end{equation}
 in the intermediate range. This scaling, confirmed by numerical data (Fig. \ref{f:R1}c), can be extended to $Fr=1$; then, $\tau_{tr}=\tau_{NL}$: the energy is transferred to small scales in an eddy turn-over time, the hallmark of FDT. }

{Note that a second characteristic  time was also introduced in \citet{mater_14}, based again on  $\epsilon_V$, and now on viscosity, namely:
\begin{equation}
T_\lambda=[\nu/\epsilon_V]^{1/2}=\tau_{NL}/[\beta Re]^{1/2} \  . 
\label{tlambda} \end{equation}
 The dependence of $T_\lambda$ on$\sqrt{\nu}$ indicates that this time is linked to the Taylor scale 
 $\lambda_V=u_{rms}/ \omega_{rms}=\sqrt{u_{rms}^2\nu/\epsilon_V}$, by writing $\lambda_V=u_{rms} T_\lambda$.
 Other relevant  scales are discussed in  \citet{barry_01, davis_11, mater_14}. }   

\subsection{Appendix B: A note on dimensional analysis} \label{SS:beta0}

{Control parameters can be defined using the characteristic times mentioned above:}
 \begin{eqnarray} 
R_g  \equiv  [\tau_{tr}/T_\lambda]^2  &=& \ u_{rms}^4/[\nu \epsilon_V]  \ = \ \beta^{-1}Re  \  ,  \label{new2a}\\ 
F_g \equiv \tau_w/\tau_{tr} \ \ \ &=& \ [NT_L]^{-1} \ \ \ \ =  \ \beta Fr  \ ,  \label{new2b} \\ 
 {\cal R}_I  \equiv R_g F_g^{2} \ \ \ &=&  \epsilon_V/[\nu N^2] \  \ \ \ = \ \beta {\cal R}_B \ , \label{new2c}  \end{eqnarray} 
{with ${\cal R}_B\equiv Re Fr^2$ already defined in \S \ref{S:EQ}.
 The difference between the two formulations in terms of $[Re, Fr]$ and $[R_g, F_g]$ is the appearance of the measured efficiency of energy dissipation in turbulent flows interacting with waves in the latter case as opposed to a purely dimensional expression in the former case. $F_g$ and $R_g$ correspond to the choice of definition of Froude and Reynolds numbers in \citet{maffioli_16l} in terms of 
$\epsilon_V$ (with $\epsilon_V/[N u_{rms}^2]=F_g$). Note that extending  this second set of parameters to the rotating case, one will define $Ro_g$ as $[f\tau_{tr}]^{-1}$ and thus $N/f=Ro/Fr=Ro_g/F_g$ will remain the same in both formulations. It is not clear whether considering these different parameters and characteristic scales allows for a better assessment of these flows. For example,  when taking this second set of parameters, $\tau_{tr}$ varies by a factor 20 within the confine of the data base in Table \ref{tab1}, whereas, as noted in \S \ref{S:STRUC}, $u_{rms}$ and $L_{int}$, and thus $\tau_{NL}$,  vary by of the order of a factor of $2$. }

\begin{figure*}
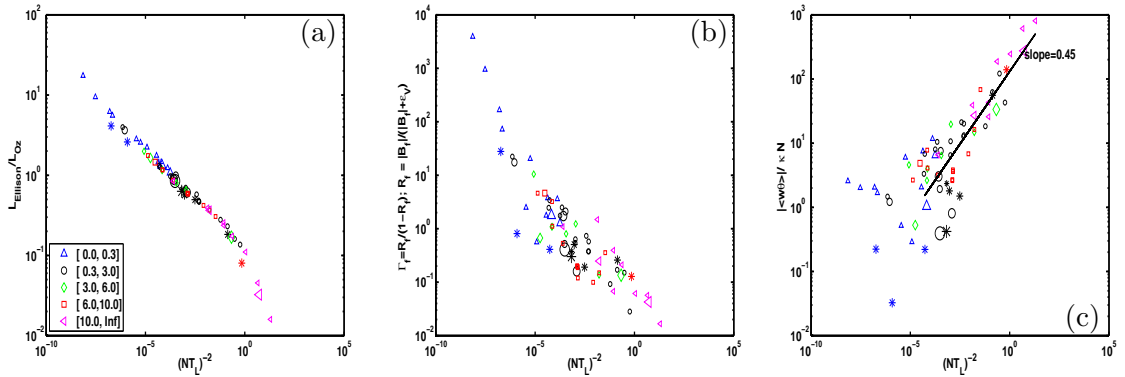
   \vspace{-1.95truecm}
\hspace{-1.1cm} \begin{minipage}{0.27\textwidth}
\large \begin{lpic}[]{pouquet-figure-9a(0.25,0.34)}
\lbl[l]{167,197;(a)} \normalsize \end{lpic} \end{minipage} 
\hspace{1.4cm}\begin{minipage}{0.27\textwidth}
\large \begin{lpic}[]{pouquet-figure-9b(0.25,0.34)}
\lbl[l]{167,197;(b)} \normalsize   \end{lpic}  \end{minipage} \vspace{-1.95truecm}
\hspace{1.1cm} \begin{minipage}{0.27\textwidth}
\large \begin{lpic}[]{pouquet-figure-9c(0.25,0.34)}
\lbl[l]{167,87;(c)} \normalsize \end{lpic} \end{minipage}
\caption{(Color online) 
Variation with $[NT_L]^{-2} =F_g^2$ of $L_{Ell}/L_{Oz}$   ({\it a}) and  of $\Gamma_f$ ({\it b}), as well as of the normalized eddy diffusivity $\kappa_\rho/\kappa$ ({\it c}), all  with binning in $Ro$
{(symbols  as in Fig. \ref{f:R1}).}
} \label{f:NT} \end{figure*}

{To illustrate this point, we examine in  Fig. \ref{f:NT} the variation with  $[NT_L]^{-2}= F_g^{+2}$  
of $L_{Ell}/\ell_{Oz}$ (a), of $\Gamma_f$ (b), and of the normalized diffusivity $\kappa_\rho/\kappa$  (c), with 
$\kappa_\rho=\left< w\theta \right>N$. The choice of power of $F_g$ on the abscissa is to be able to compare with variations in ${\cal R}_B, {\cal R}_I$. 
The least-square fit for $L_{Ell}/\ell_{Oz}\approx F_g^{-1/2}$ for regimes I and II is in agreement with the data in Fig. \ref{f:LE}, with in the intermediate regime (II), $L_{Ell}/\ell_{Oz}\approx Fr^{-1}$, and $F_g\sim Fr^2$. However, it extends now through two regimes, showing that the $F_g$ parameter allows to cross smoothly through these regimes, with only saturation when $F_g\gtrsim 1, Fr\gtrsim 1$, now moving into the third regime of stratified but strong  turbulence. \\       
The two scalings that can be identified for the mixing efficiency $\Gamma_f$ are in agreement with previous figures, in particular with $\Gamma_f\sim F_g^{-2}$ at small $NT_L$, and $\Gamma_f\sim F_g^{-1}$ at large $NT_L$, with a cross-over at $F_g^2 \approx 10^{-6}$. \\
Finally, the normalized eddy diffusivity  again shows  scaling behavior for $F_g$ sufficiently large. At lower $F_g$, it saturates to values close to unity except for three data points which are both at low Froude and low Rossby numbers, with Richardson numbers between 5 and 10 and buoyancy Reynolds numbers close to 5: these flows are in a transitional regime sensitive to fluctuations close to the threshold of instabilities. }

\subsection{Appendix C: Derived dimensionless parameters} \label{SS:C}

{One {can} also define micro-Froude and micro-Rossby numbers, $F_\omega$ and $R_\omega$, based on the effective kinetic energy dissipation rate $\epsilon_V$, with $\omega_{rms}$ the {\it rms} vorticity, $\mathbf{\omega}=\nabla \times {\bf u}$:
  \begin{equation}
F_\omega \equiv \bigg[\frac{\epsilon_V}{\nu N^2}\bigg]^{1/2} =\frac{\omega_{rms}}{N}   \ , \ 
R_\omega  \equiv \bigg[\frac{\epsilon_V}{\nu f^2}\bigg]^{1/2} =\frac{\omega_{rms}}{f}    \ .
\label{micro0}   \end{equation}
The runs of Table \ref{tab1} have $11.5 \le R_\omega \le 3244$: the intrinsic vorticity of the flow dominates the imposed rotation at small scales for all runs.
Note that  
$${\cal R}_I  \ \equiv \ F_\omega^2 \ = \epsilon_V/[\nu N^2]$$
 is called any of: the buoyancy Reynolds number \citep{ivey_08},  or the activity parameter \citep{stretch_10b}, or the turbulence intensity parameter \citep{delavergne_16}. The buoyancy Reynolds number ${\cal R}_B$ is what ${\cal R}_I$ would be under the assumption that the turbulence has reached its full potential, and that the dissipation rate is equal to its dimensional expression, $\epsilon_D$. 
Thus, ${\cal R}_I$ is an expression that is compatible with a small-scale Kolmogorov isotropic energy spectrum, with ${\cal R}_B$ and ${\cal R}_I$ differing by a factor $\beta$, namely 
${\cal R}_B \equiv Re Fr^2 = \epsilon_{{\bf D}}/[\nu N^2]  = \beta^{-1} {\cal R}_I$. In terms of ratio of characteristic length scales,  assuming a Kolmogorov spectrum, $E_V(k)\sim \epsilon_V^{2/3} k^{-5/3}$, one can also write:
  \begin{equation}
{\cal R}_I  = \  [\ell_{Oz}/\eta]^{4/3}  \  ,  \ \  R_\omega  = \ [\ell_{Ze}/\eta]^{2/3}  \ = \ \beta^{1/2} Re^{1/2} Ro     \ , 
 \label{micro1} \end{equation}
with as usual the dissipation and Ozmidov scales defined as  $\eta = 2\pi  [\epsilon_V/\nu^3]^{-1/4}, \ell_{Oz} =2\pi  \sqrt{\epsilon_V/N^3}$ (see   \S \ref{S:EQ}), and equivalently for rotation the Zeman scale $ \ell_{Ze}=2\pi \sqrt{\epsilon_V/f^3}$.
It is thus clear that ${\cal R}_I$ represents a dimensional estimate of the ratio of inertial to dissipative forces for stratified turbulence. }
\vskip0.15truein

\bibliographystyle{jfm}  \bibliography{ap_17_nov03-copy}  

\end{document}